# Theory of surface plasmon generation at nanoslit apertures

P. Lalanne, J.P. Hugonin and J.C. Rodier

Laboratoire Charles Fabry de l'Institut d'Optique, Centre National de la Recherche Scientifique, F-91403 Orsay cedex, France

**Abstract:**

In this letter, we study the scattering of light by a single subwavelength slit in a metal screen. In contrast to previous theoretical works, we provide a microscopic description of the scattering process by emphasizing the generation of surface plasmons at the slit apertures. The analysis is supported by a rigorous formalism based on a normal-mode-decomposition technique and by a semi-analytical model which provides accurate formulae for the plasmonic generation strengths. The generation is shown to be fairly efficient for metals with a low conductivity, like gold in the visible regime. Verification of the theory is also shown by comparison with recent experimental data [H.F. Schouten et al., Phys. Rev. Lett. **94**, 053901 (2005)].

PACS :

42.25.Fx Diffraction and scattering

42.79.Ag Apertures, collimators

73.20.Mf Collective excitations

78.66.-w Optical properties of specific thin films

Light scattering by a single subwavelength aperture in a metal screen, like a hole or a slit, represents a basic diffraction phenomenon which has been studied for a long time [1]. Following the discovery of the extraordinary transmission [2] through nanohole arrays in thick metallic films, this basic diffraction problem has recently sparked a wealth of experimental and theoretical works aiming at understanding the underlying physics and at developing potential applications of light manipulation on a subwavelength scale [3]. Single nanoapertures, isolated or dressed by shallow corrugation [4-5-6-7], are currently investigated at visible and microwave frequencies. Studies of the aforementioned structures have revealed that surface plasmon polaritons (SPPs) launched at the input and output sides of the nanoaperture drastically impact the scattering properties. The SPP interpretation has been evidenced by recent experimental works revealing surprising phenomena, like the strong angular confinement of light transmitted through a single nanoaperture dressed by shallow corrugation [4] or the modulation of the total far-field intensity diffracted by a pair of slits as in Young's experiment [8], or by direct near-field measurements [9]. On a theoretical side, little is know on the SPP generation at the nanoaperture entrance. Earlier works [1] have considered perfectly-conducting metal and have not addressed the problem. Whether they rely on semi-analytic [10,5,6] or on intensive computational approaches [9,11-12], recent works well support far-field experimental data on the transmission for instance, but do not quantitatively address the SPP generation. Calculated near-fields qualitatively support the existence of SPPs, but no microscopic mechanism for the coupling to SPPs is provided : for instance, the question arises how much light is scattered into SPPs at a nanoscale aperture entrance.

In this letter, we consider the SPP generation at a metallic interface perforated by an isolated subwavelength slit, when the slit is illuminated by its fundamental guided mode (Fig. 1a) or by an incident plane wave (Fig.1b). The configuration is semi-infinite on the two - z-half spaces. We first present a general theoretical formalism for the calculation of the scattering coefficients between the incident light and the SPP modes launched at the slit aperture. Rigorous calculations show that, for metals with a low conductivity like gold in the visible regime, the fraction of the incident radiation which couples into the SPP is fairly large, reaching values as high as the total far-field radiated energy. In a second step, we derive a semi-analytical model which quantitatively agrees with the rigorous calculations. This derivation is motivated by the fact that simple intuitive theoretical formalisms have not been yet presented for the SPP generation at apertures, despite its importance for light manipulation at a subwavelength scale. Realistic expressions for the generation strength may be also useful for experimentalists in their plasmonic devices or instruments. The model allows us to understand the influence of the different parameters of the scattering problem. Additionally, it quantitatively interprets the recent observation of far-field intensity modulations in the double-slit experiment [8].

Hereafter, the metal is considered as a real metal with a finite conductivity. Gold will be used to illustrate our discussion and its frequency-dependent permittivity values $\varepsilon$ is taken from Ref. [13]. Let us first consider the geometry shown in Fig. 1a. The slit has the same direction as the magnetic field $H_y$ (transverse-magnetic polarization) and is illuminated by its fundamental guided mode at a fixed wavelength $\lambda$ ($k = 2\pi/\lambda = \omega/c$). In the figure, $n_1$ and $n_2$ refer to the refractive indices of the dielectric materials inside and below the slit. We denote by $w$ the slit width. Inside the slit, the magnetic field admits a modal expansion of the form

$$H_1(x,z) = \Psi_0(x) \exp(ik\, n_0^{eff}\, z) + \Sigma_p\, r_p\, \Psi_p(x) \exp(ik\, n_p^{eff}\, z), \qquad (1)$$

where p is an integer, $\Psi_p$ is the magnetic field of mode p and $n_p^{eff}$ its normalized propagation constant. Similarly below the slit, the magnetic field $H_2$ can be expanded onto a plane wave basis

$$H_2(x,z) = \int_{-\infty}^{\infty} du\, t_u \exp(ikn_2ux)\exp(ikn_2\gamma_u z), \qquad (2)$$

where $u^2+(\gamma_u)^2 = 1$. Many numerical tools can provide electromagnetic-field distributions inside and below the slit by rigorously solving for the reflection $r_p$ and transmission $t_u$ amplitudes [8-12]. Here, we have used a Fourier modal method [14] for that purpose. For visible light, $\lambda = 0.6$ μm, Fig. 2a shows the electromagnetic fields $H_y(x,z)$ and $E_z(x,z)$ which will be of main concern in the following. The near-field pattern below the slit is a superposition of many evanescent and propagative waves and exhibits a marked expansion of several wavelengths away from the slit aperture in the x-direction. This is consistent with the involvement of SPPs at the slit aperture. However, the respective roles of the evanescent field and of the SPP waves deserve to be clarified. To explore this question, we propose an original approach which basically relies on the completeness theorem for the normal modes of waveguides [15]. That theorem, which provides a useful electromagnetic representation of light propagation in translationally invariant systems, stipulates that any transverse field pattern of such a system can be decomposed as a linear combination of forward- and backward-travelling bounded and radiative modes. Thus for our slit geometry and for $x > |w/2|$, it results that the transverse electromagnetic fields shown in Figs. 2a-2b can be expanded into the set of normal modes of a flat gold-air interface [16]:

$$H_y = \left(\alpha^+(x) + \alpha^-(x)\right) H_{SP}(z) + \sum_\sigma a_\sigma(x) H_\sigma^{(rad)}(z), \qquad (3a)$$

$$E_z = \left(\alpha^+(x) - \alpha^-(x)\right) E_{SP}(z) + \sum_\sigma a_\sigma(x) E_\sigma^{(rad)}(z). \qquad (3b)$$

In Eqs. 3a-3b, the summation refers to radiative modes and $\{H_{SP}, E_{SP}\}$ are the transverse magnetic and electric fields of a SPP mode with unit intensity launched at a metallic/dielectric interface. These fields are known analytically [17] : $H_{SP}(z) = (N_{SP})^{-1/2} \exp(i\gamma_{SP} z)$, with $N_{SP}$ a normalization constant and $\gamma_{SP} = [\varepsilon k^2 - (k_{SP})^2]^{1/2}$ in the metal and $[(n_2 k)^2 - (k_{SP})^2]^{1/2}$ in the dielectric, respectively. $k_{SP} = k\left[\left(\varepsilon n_2^2\right)/\left(\varepsilon+n_2^2\right)\right]^{1/2}$ is the propagation constant of the SPP. In Eqs. 3a-b, $\alpha^+(x)$ and $\alpha^-(x)$ are coefficients of central importance to the present discussion related to the SPP generation. The plus and minus superscripts refer to SPP propagating forward and backward. The x-dependence is known analytically : $\alpha^+(x) = \alpha^+(w/2) \exp(i\, k_{SP}\, (x-w/2))$ and $\alpha^-(x) = \alpha^-(w/2) \exp(-i\, k_{SP}\, (x+w/2))$, where $\alpha^+(w/2)$ and $\alpha^-(-w/2)$ represent the unknown complex coefficients related to the strengths of the SPP excitation at the exit sides of the slit. Because of the mode orthogonality condition, we have [16]:

$$\int_{-\infty}^{\infty} dz\, H_y(x,z)\, E_{SP}(z) = 2\left(\alpha^+(x) + \alpha^-(x)\right), \tag{4a}$$

$$\int_{-\infty}^{\infty} dz\, E_z(x,z)\, H_{SP}(z) = 2\left(\alpha^+(x) - \alpha^-(x)\right). \tag{4b}$$

From Eqs. 4a-4b, the coefficients $\alpha^+(x)$ and $\alpha^-(x)$ can be derived by calculating numerically the overlap integrals on the left-hand sides of the equations. Figure 2b represents the modulus squared of these coefficients for the scattering problem considered in Fig. 2a. The results are obtained for an incident slit mode with a unit intensity ; thus $|\alpha^+(x)|^2$ and $|\alpha^-(x)|^2$ represent the normalized SPP generation efficiencies. For $-w/2 < x < w/2$, the coefficients are meaningless since there is no air-metal interface. Let us consider $|\alpha^+(x)|^2$, the discussion is similar for $|\alpha^-(x)|^2$. For $x > w/2$, the computed $|\alpha^+(x)|^2$ values perfectly fulfil the expected SPP attenuation law, $|\alpha^+(w/2)|^2 \exp[-2\mathrm{Im}(k_{SP})(x-w/2)]$ shown as circles in Fig. 2b. Further analysis, not reported here, have shown that the phase dependence of $\alpha^+(x)$ exactly coincides with the SPP propagation constant $\exp[-i\mathrm{Re}(k_{SP})(x-w/2)]$. Consistently with the outgoing radiation condition at the slit output, $|\alpha^+(x)|^2$ is found to be approximately null for $x < w/2$, $|\alpha^+(x)|^2 < 10^{-6}$ for any x in the computational window. In our opinion, all this provides a strong support for the soundness of the normal-mode-decomposition formalism used for quantifying the SPP excitation.

In order to gain better physical insight into the problem and to obtain useful expressions for the SPP generation, we have developed an intuitive approximate model. As suggested by the rigorous formalism, the SPP generation results from a two-stage mechanism : a purely geometric diffraction problem followed by the launching of a SPP bounded mode on a flat interface. We have assumed that the diffraction problem which results in a specific near-field distribution in the immediate vicinity of the slit (the $H_y, E_z$ fields of Eqs. 4a-b at $x = \pm w/2$) is weakly dependent on the dielectric properties of the metal and that it can be estimated by considering the metal as a perfect conductor. On the contrary, the launching of the bounded SSP mode strongly depends on the intrinsic dielectric properties of the metal-dielectric interface (the $H_{SP}, E_{SP}$ fields of Eqs. 4a-b). Assuming an ideal metal drastically reduces the scattering problem complexity and, under the approximation that the field in the slit is solely composed of the forward-incident and backward-reflected fundamental modes, the field distribution $\{H_y(\pm w/2,z), E_z(\pm w/2,z)\}$ can be calculated analytically. Our approach follows that developed in [18] for slit arrays. From the near-field distribution obtained for a slit in a perfect metal, the overlap integrals of Eq. 4a-b is subsequently calculated for $x = \pm w/2$. We do not provide here the details of the lengthy calculation and just give the results. One obtains :

$$\alpha^+(w/2) = \alpha^-(-w/2) = \frac{1}{i}\left(\frac{4}{\pi}\frac{n_2^2}{n_1}\frac{\sqrt{|\varepsilon|}}{-\varepsilon-n_2^2}w'\right)^{1/2}\frac{I_1}{1+(n_2/n_1)w'I_0}, \qquad (5)$$

where $w' = n_2 w/\lambda$ is the normalized slit width, $u^2 + (\gamma_u)^2 = 1$, $I_0 = \int_{-\infty}^{\infty} du\, \mathrm{sinc}^2(\pi w'u)/\gamma_u$ and $I_1 = \int_{-\infty}^{\infty} du\, \frac{\mathrm{sinc}(\pi w'u)\cos(\pi w'u)}{\gamma_u\left(\gamma_u + \sqrt{n_2^2/(\varepsilon+n_2^2)}\right)}$ which are easily calculated with classical integration techniques. Note that $I_0$ is independent of the metal permittivity and that $I_1$ is only weakly dependent. Similarly, for the geometry depicted in Fig. 1b and for an incident plane-wave with a unitary integrated power on the opening, we found :

$$\beta^+(w/2) = \beta^-(-w/2) = -\sqrt{\frac{n_2}{n_1}} \frac{\text{sinc}(\pi w'\theta)}{\sqrt{\cos(\theta)}} \alpha^+(w/2), \qquad (6)$$

for small incidence angles.

Figure 3 summarizes the main useful results predicted for slits in gold over a broad spectral interval ranging from the visible to thermal infrared. All plots are relative to the total SPP-excitation efficiency $e_{SP}$ on both sides of the aperture, $e_{SP}$ being equal to $|\alpha^+(w/2)|^2+|\alpha^-(w/2)|^2$ in (a) and (b), and to $|\beta^+(w/2)|^2+|\beta^-(w/2)|^2$ in (c). In order to access the accuracy of the model predictions (solid curves), we have performed extensive calculations using the theoretical formalism, as in Fig. 2. The corresponding data are shown with marks. A quantitative agreement is obtained for all cases ; the larger the wavelength, the better the agreement as expected from the perfectly-conducting metal approximation. At visible frequencies, the SPP excitation is surprisingly efficient. For an optimal slit width of $w \approx 0.23\lambda$ (a value nearly independent of the metal permittivity), as much light (≈40%) is scattered into the SPPs as it is radiated in the far-field. For longer wavelengths, $e_{SP}$ rapidly decreases ; the maximum efficiency is only 2.8% for $\lambda = 10$ μm, and is predicted to be as low as 1% for $\lambda = 18$ μm. From Eq. (5), it is easily shown that the efficiency scales as $|\varepsilon(\lambda)|^{-1/2}$ because $|\varepsilon| << (n_2)^2$. Figure 3b shows the impact of the substrate refractive index. As $n_2$ increases, $e_{SP}$ increases as well ($|\alpha^+(w/2)|^2 \propto n_2/n_1$), and the optimal slit width for best excitation slightly decreases. As shown in Fig. 3c, these trends remain valid when the slit aperture is illuminated by an incident plane-wave, a 50%-efficient SPP-excitation being predicted for $\lambda = 0.6$ μm and for the optimal slit width. This indicates that even simple metallic structures, like slits, may be used to efficiently manipulate SPP visible waves at a nanoscale level.

To further validate the model, we now rub it with experimental data. The recent plasmon-assisted Young's type experiment [8] provides a test-bed for our microscopic theoretical treatment. In this experiment, the total power radiated by two slits separated by a few micrometers has been shown to be enhanced or reduced as a function of the wavelength of the incident beam. The transmission through the slits can be predicted by use of a classical Fabry-Perot model [18] by assuming that the energy transfer within the slit is solely performed through the fundamental slit mode. However, calculations have shown that the bouncing of the fundamental mode marginally affects the analysis and that the experiment data can be interpreted by simply evaluating the power S coupled into each single slit, see Fig. 4a. Reciprocity warrants that the scattering coefficient from SPP to slit-mode is equal to the $\alpha^-$, which corresponds to the back-conversion process from slit-mode to SPP. Therefore, for an incident plane wave with a unit power per slit-opening area, S is given by $S = |t_0 + \alpha^- \beta^+ \exp[ik_{SP}(d-w)]|^2$. The modulation of the coupled power S is shown in Figs. 4b-4c. The contrast $C = 2|t_0/(\alpha^- \beta^+)| \approx 0.26$ which is nearly constant in the spectral range well agrees with the experimental values $C \approx 0.2$, but more importantly, the minima and maxima of S exactly coincide with the experimental ones, see Fig. 1 in [8] for comparison, evidencing that the important phase factors associated to the SPP scattering and back conversion processes is well predicted by the model.

In conclusion, we have studied the generation of SPP waves at a metallic interface perforated by an isolated subwavelength slit. A rigorous formalism based on a normal-mode-decomposition technique has been derived. We have also proposed a semi-analytical model which has been validated by comparisons with computational data obtained from the rigorous formalism and with experimental data recently reported for the transmission through a slit doublet. It has been shown that the SPP-generation processes can be very efficient at visible frequencies for noble metals. The rigorous formalism which relies on fundamentals in electromagnetism can be applied to other related geometries and we believe that the semi-analytical model could be as well generalized to slit arrays or holes. Thus the theoretical treatment is expected to be able to provide a microscopic description of various plasmonic devices of current interest.

This work is partly supported by the Network of Excellence on Micro Optics NEMO.

# Figure Captions

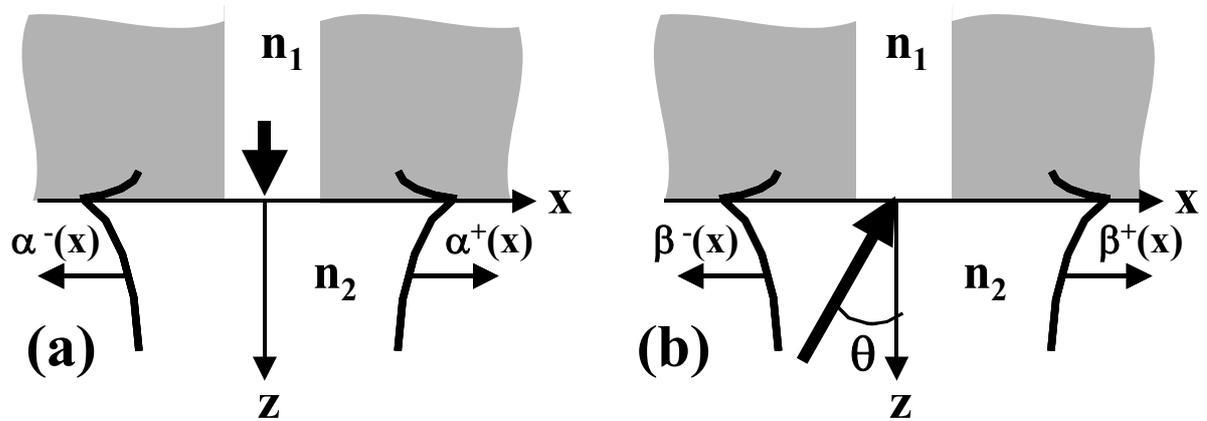

**Figure 1**

SPP generation at a metallic interface perforated by a single slit under illumination by the fundamental mode of the slit (a), or by an plane wave with an incidence angle θ (b). The slit width is denoted by $w$ and $n_1$ and $n_2$ refer to the refractive indices inside and bellow the slit. $\alpha^+(x)$, $\alpha^-(x)$, $\beta^+(x)$ and $\beta^-(x)$ are the SPP-generation coefficients.

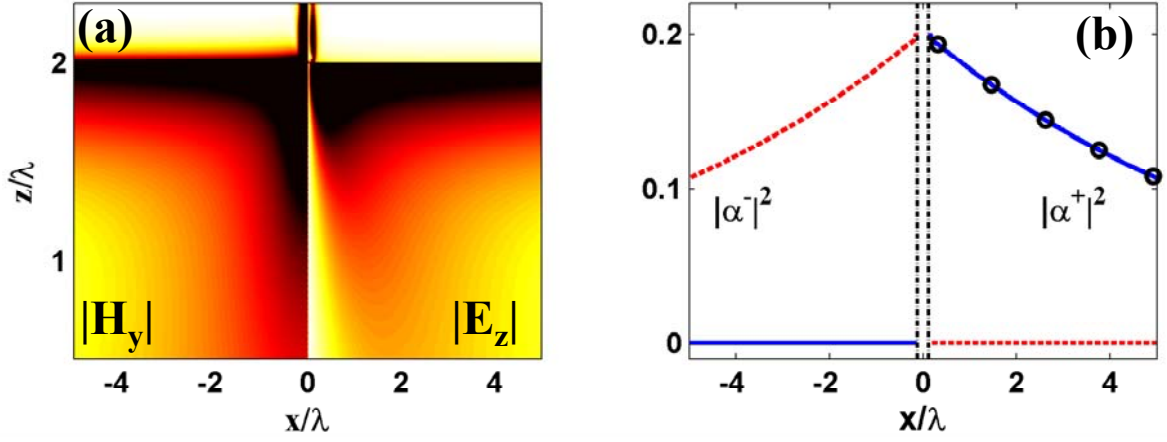

**Figure 2**

(color online). Validation of the theoretical formalism for the calculation of the SPP-generation coefficients. (a) : Near-field pattern generated at a gold interface by the scattering of the fundamental guided mode in the slit. The left hand-side refers to $|H_y|$ and the right-hand side to $|E_z|$. A saturated scale is used for reinforcing the fields in the vicinity of the slit. (b) : SPP generation strengths, $|\alpha^+(x)|^2$ (solid) and $|\alpha^-(x)|^2$ (dotted), obtained from the fields in (a) by calculating the overlap integrals of Eqs. 4a-b. The circles are numerical data equal to 0.202 exp[-2Im($k_{SP}$) (x-$w$/2)], showing that the exponential decay of $|\alpha^+(x)|^2$ results from the expected SPP damping. Parameters are $\lambda$ =0.6 µm, $w$ = 0.14 µm and $n_1 = n_2 = 1$.

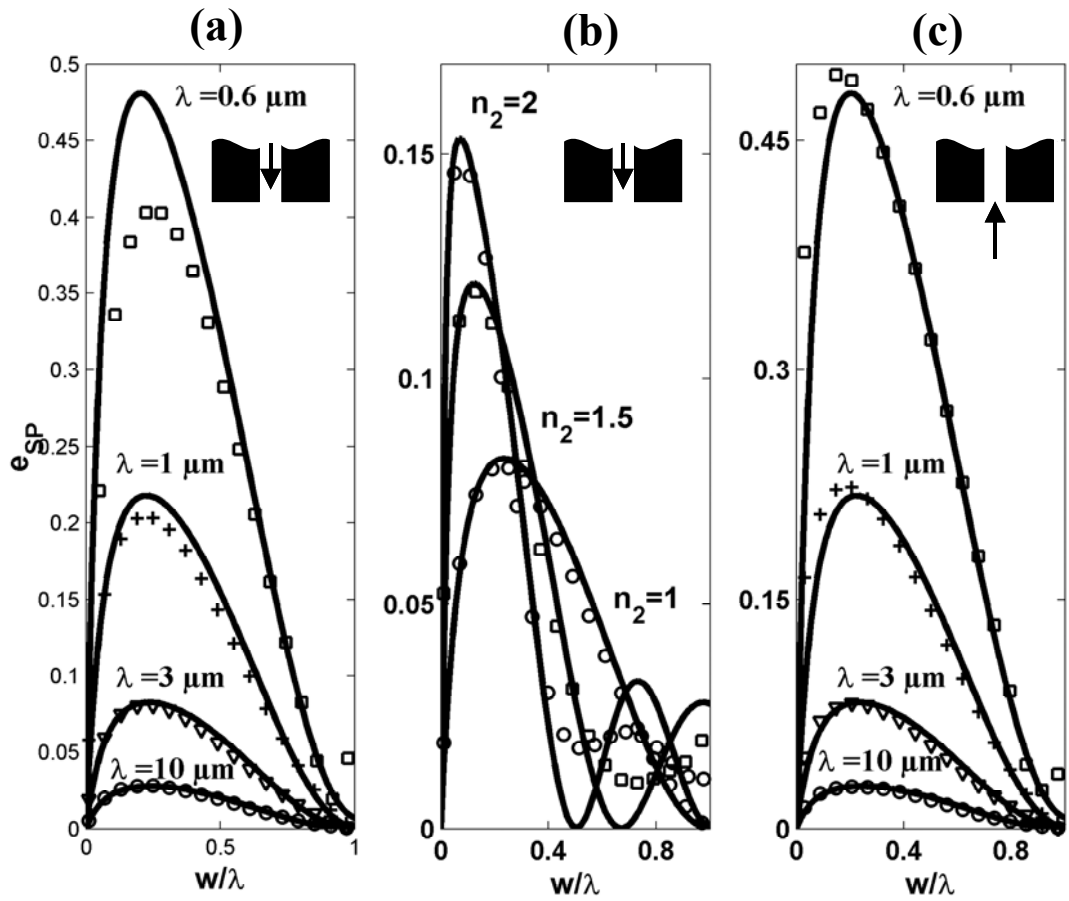

**Figure 3**

Total SPP-generation efficiencies $e_{SP}$ as a function of the slit width for different geometries of practical interest. Note the change of vertical scales. (a) Illumination by the fundamental mode of the slit for different wavelengths, $n_1 = n_2 = 1$. (b) Influence of the substrate refractive index $n_2$, $n_1 = 1$ and $\lambda = 3$ µm. (c) Illumination by the incident plane wave for different wavelengths, $n_1 = n_2 = 1$ and $\theta = 0$. Marks (circles, squares …) are calculated data obtained with the rigorous formalism. Solid lines represent the model predictions. The incident waves have an unitary integrated power on the opening.

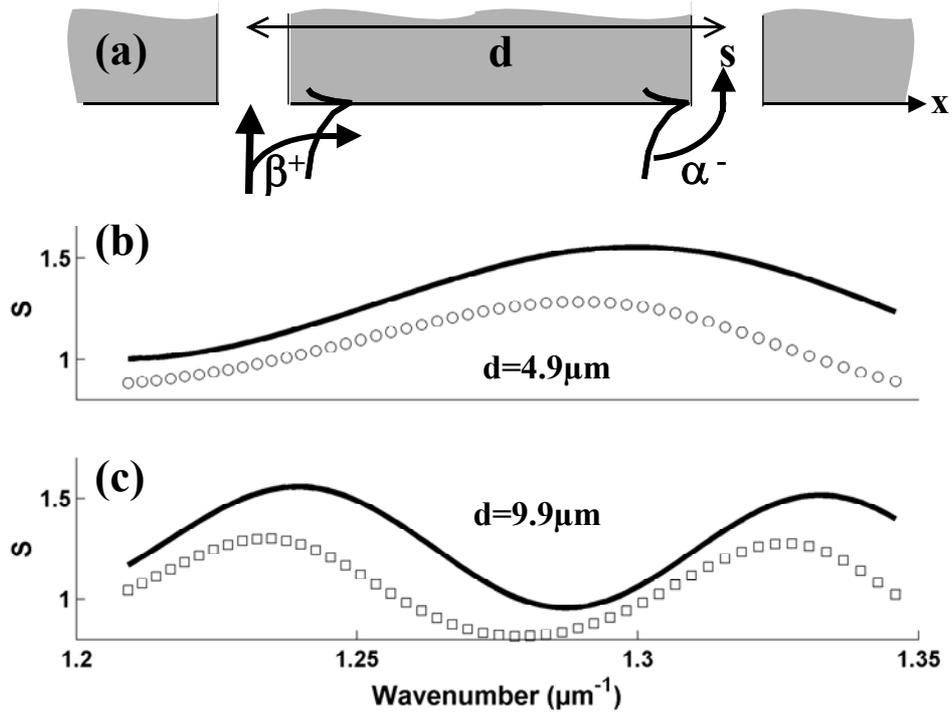

**Figure 4**

Model validation against experimental data . (a) double-slit Young's experiment as in [8]. (b) and (c) : Model prediction for d = 4.9 μm and 9.9 μm. Since the horizontal axis is the same as in Fig. 1 in [8], a direct visual comparison is allowed with experimental data. Circles and squares represent computational data obtained with the Fourier modal method [14]. These rigorous data are only slightly shifted with respect to the approximate model predictions.

[16] In Eqs. 3a-3b, the bounded SPP mode is the analogue of the guided mode of the waveguide theory, while the summation represents a continuum of radiation modes of the flat metal/dielectric interface. Because we are concerned by metals with a finite conductivity, all

these modes are not orthogonal in the sense of the Poynting-vector as it is usually the case in waveguide theory with lossless materials, but they obey the unconjugate general form of orthogonality, see A.W. Snyder and J.D. Love, *Optical Waveguide theory*, (Chapman and Hall, NY, 1983) for more details. That is the reason why EH products (instead of the usual EH* product) are used in Eqs. 4a-b.

[17] H. Raether, *Surface Plasmons on Smooth and Rough Surfaces and on Gratings,* (Springer-Verlag, Berlin, 1988).

[18] Ph. Lalanne et al., J. Opt. A: Pure Appl. Opt. **2**, 48 (2000).